\documentclass[twocolumn,preprintnumbers,amsmath,amssymb]{revtex4}

\usepackage{color}
\usepackage{graphicx}
\usepackage{dcolumn}
\usepackage{bm}
\usepackage{comment}
\usepackage{float}

\def\lsim{\mathrel{\rlap{\lower4pt\hbox{\hskip1pt$\sim$}}
\raise1pt\hbox{$<$}}}
\def\gsim{\mathrel{\rlap{\lower4pt\hbox{\hskip1pt$\sim$}}
\raise1pt\hbox{$>$}}}

\def\draftmark {
}

\begin{document}
\draftmark

\title{Null-stream pointing with pulsar timing arrays}

\author{Jeffrey S.\ Hazboun} \email{jeffrey.hazboun@gmail.com}
\affiliation{Department of Physics, Hendrix College, Conway, AR
72032}

\author{Shane L.\ Larson} \email{s.larson@northwestern.edu}
\affiliation{Center for Interdisciplinary Exploration and Research in
Astrophysics, Northwestern University, Evanston, IL 60208\\
	     and Department of Astronomy, Adler Planetarium, Chicago,
	     IL 60605}

\date{\today}

\begin{abstract}
Locating sources on the sky is one of the largest challenges in
gravitational wave astronomy, owing to the omni-directional nature of
gravitational wave detection techniques, and the often intrinsically
weak signals being observed.  Ground-based detectors can address the
pointing problem by observing with a network of detectors, effectively
triangulating signal locations by observing the arrival times across
the network.  Space-based detectors will observe long-lived sources
that persist while the detector moves relative to their location on
the sky, using Doppler shifts of the signal to locate the sky
position.  While these methods improve the pointing capability of a
detector or network, the angular resolution is still coarse compared
to the standards one expects from electromagnetic astronomy.  Another
technique that can be used for sky localization is null-stream
pointing.  In the case where multiple independent data streams exist,
a single astrophysical source of gravitational waves will appear in
each of the data streams.  Taking the signals from multiple detectors
in linear combination with each other, one finds there is a two
parameter family of coefficients that effectively null the
gravitational wave signal; those two parameters are the angles that
define the sky location of the source.  This technique has been
demonstrated for a network of ground-based interferometric
observatories, and for 6-link space interferometers.  This paper
derives and extends the null-stream pointing method to the unique case
of pulsar timing residuals.  The basic method is derived and
demonstrated, and the necessity of using the method with multiple
sub-arrays of pulsars in the pulsar timing array network is
considered.
\end{abstract}


\maketitle

\section{Introduction}\label{sec.intro}
The gravitational wave spectrum covers many decades in frequency
space, just like the electromagnetic spectrum.  The particular waves
that are radiated in any given band of the spectrum reflect the
astrophysical phenomena that generated the waves, and in particular
the astrophysical timescales that dominate the movement of mass in the
system.  In the very-low frequency band of the spectrum, $10^{-9} Hz
\lesssim f_{gw} \lesssim 10^{-6} Hz$, the primary detection technique
is known as pulsar timing.

Pulsar timing uses the stable rotation of distant pulsars as clocks.
It was first described by Detweiler \cite{Detweiler1979}, and proceeds
as follows.  The arrival time of pulses from a pulsar are monitored on
Earth, and compared against a model for the expected arrival times
(the ``ephemeris'').  Using simple models for pulsar spindown over
time, models for pulse arrival times can be built with precisions at
the level of fractions of a microsecond when pulsar monitoring spans
several years.  The fundamental signature of a gravitational wave
passing between the Earth and the distant pulsar is a change in the time
of flight for individual pulses, advancing or retarding their time of
arrival compared to the model ephemeris.  The difference between the
arrival time of the pulses and the model are known as ``timing
residuals,'' and they constitute the basic data stream in pulsar
timing searches for gravitational waves.

While the measurement can be made with long-term observations of a
single pulsar, the implementation of this method as a viable detection
technique has been realized through the development of \textit{pulsar
timing arrays}, where many pulsars in many parts of the sky are
monitored over long periods of time, combining timing residuals from
multiple pulsars to search for gravitational waves.  Many efforts have
been launched on this front, including the North American Nanohertz
Observatory for Gravitational Waves (NANOGrav)\cite{nanograv}, the
Parkes Pulsar Timing Array \cite{PPTA}, the European Pulsar Timing
Array \cite{EPTA}, and the International Pulsar Timing Array
\cite{IPTA}.

The expected astrophysical sources of very-low frequency gravitational
waves include supermassive black hole binaries (SMBHB), stochastic
backgrounds of SMBHBs, as well as a variety of other stochastic 
sources such as phase transitions in the early Universe or relic 
radiation from the Big Bang. See \cite{Burke-Spolaor:2015xpf} for a recent review of PTA sources. 

Like most gravitational wave detection techniques, pulsar timing
arrays are omnidirectional, in the sense that they are sensitive to
gravitational waves from any location on the sky.  As a general rule
of thumb, the sensitivity of any particular pulsar to incident
gravitational waves is a function of the angle between the line of
sight to the pulsar and the line of sight to the gravitational wave
source. This relationship is distinctly evident in the Hellings and Downs curve \cite{Hellings:1983fr}, which relates the correlation of residual signals from two pulsars to their angular separation in the sky. 

This work develops a data combination technique known as
``pulsar null streams'' to provide a good estimate of the location of
a gravitational wave source on the sky.  This is an explicit analysis
technique for determining the sky location of a putative gravitational
wave source without engaging in a full parameter search.  Source
location knowledge absent other parameter information is useful for
counterpart searches, as well as restricting the search space of 
computationally intensive signal searches.

Null stream mapping of gravitational wave sources has been described
for interferometric detectors \cite{TintoGursel,ZSS1,ZSS2,NullNetwork},
and relies on the fact that there are correlated gravitational wave
signals between detectors in a network.  For the case of a pulsar
timing array, one has the same situation --- a gravitational wavefront
will produce a correlated response in the timing of every pulsar in
the array.  This correlation may be exploited to create a null stream
by taking advantage of the geometrical properties of a pulsar's
response to incident gravitational waves.

The paper is organized as follows.  In section \ref{sec.pulsarTiming}
we review a basic signal model for pulsar timing residuals, and
express it in a form conducive to build a pulsar null stream.  In
section \ref{sec.pns} the pulsar null stream is described and written
out.  Section \ref{sec.demonstration} shows how the pulsar null stream
works for an array of three pulsars. 
Section \ref{sec.PulsarSep} discusses the errors inherent in one sub-array of pulsars. Section \ref{sec.MultSubArray} demonstrates how multiple sub-arrays of pulsars strengthen the null signal technique. 
Section \ref{sec.noise} examines the efficacy and
overall pointing ability of the method in the presence of noise.
Section \ref{sec.discussion} summarizes the key results and discusses
future directions for this work.

\section{Pulsar timing residuals}\label{sec.pulsarTiming}

Pulsar timing arrays use the long term stability in the spacing of
signal pulses from radio pulsars to detect and characterize
gravitational waves.  A passing gravitational wave advances or delays
the arrival time of regular pulses at the Earth from the pulsar.  The
advance or delay of the pulse is found by subtracting the pulse record
from a model of the pulse arrival times in the absence of a
gravitational wave; the result is referred to as the residual,
$R_{g}(t)$.  The residual of a pulsar signal is dependent on the sky
location and physical characteristics of the gravitational wave source
as well as the sky location and distance to the pulsar being timed.
Consider a binary source of gravitational waves located on the sky at
an ecliptic longitude $\lambda$ and an ecliptic latitude $\beta$.  For
a pulsar located at $(\lambda_{P},\beta_{P})$ the timing residual is
given by~\cite{Lee:2011et}
\begin{widetext}
\begin{align} \label{Residual}
    R_{g}\left(t\right)=&\frac{1}{2\omega_{g}}
    \frac{\sin\left(\Delta\Phi\right)}
    {1-\cos\theta}\left[\left(B_{1}\cos\left(2\phi\right)+B_{2}
    \sin\left(2\phi\right)\right)h_{+}\left(\omega_{g}t-\Delta\Phi\right)
    +\left(B_{2}\cos\left(2\phi\right)-B_{1}\sin\left(2\phi\right)\right)
    h_{\times}\left(\omega_{g}t-\Delta\Phi\right)\right]\ ,
\end{align}
\end{widetext}
where $\omega_{g} = 2\pi f$ is the gravitational wave
frequency, $\phi$ is the ascending node of the source binary orbital
plane, and the remaining terms are defined as
\begin{align}
    B_{1}=
    &\left(1+\sin^{2}\beta\right)\cos^{2}\beta_{P}\cos\left[2\left(\lambda-\lambda_{P}\right)\right]
    \\
    &
    -\sin\left(2\beta\right)\sin\left(2\beta_{P}\right)\cos
    \left(\lambda-\lambda_{P}\right)\nonumber\\
    & +\left(2-3\cos^{2}\beta_{P}\right)\cos^{2}\beta\nonumber \\
    B_{2}=
    &2\cos\beta\sin\left(2\beta_{P}\right)\sin\left(\lambda-
    \lambda_{P}\right)\\
    &-2\sin\beta\cos^{2}\beta_{P}\sin\left[2\left(\lambda-
    \lambda_{P}\right)\right]\nonumber\\
    \Delta\Phi&=\frac{1}{2}\omega_{g}D_{P}\left(1-\cos\theta\right)\\
    \cos\theta&=\cos\beta\cos\beta_{P}\cos\left(\lambda-
    \lambda_{P}\right)+\sin\beta\sin\beta_{P}
\end{align}
This form of the residual from \cite{Lee:2011et} is written so that the pulsar term is taken into account as a phase shift and amplitude modulation and can be tracked by the presence of $\Delta\Phi$, where the distance to the pulsar resides. 

The Fourier transform of the residual
can be written as $ {\tilde R}_{i}(f) = {\cal
F}_{i}^{+} {\tilde h}_{+}(f) + {\cal F}_{i}^{\times} {\tilde
h}_{\times}(f)$, where the ${\cal F}_{i}^{+,\times}$ are ``beam
pattern functions'' for the $i^{th}$ pulsar and are parameterized by the
sky position of the source, $\{\beta, \lambda\}$.  They can be read off as the coefficients of the individual polarizations of the waveform in Eq~(\ref{Residual}).
\begin{subequations}\label{BeamPattern1}
\begin{align}  
	{\cal F}_{i}^{+}=		&\frac{1}{2\omega_{g}}
    \frac{\sin\left(\Delta\Phi_{i}\right)}
    {1-\cos\theta_{i}}\left(B_{1i}\cos\left(2\phi_{i}\right)+B_{2i}
    \sin\left(2\phi_{i}\right)\right)\\
	{\cal F}_{i}^{\times}=	&\frac{1}{2\omega_{g}}
    \frac{\sin\left(\Delta\Phi_{i}\right)}
    {1-\cos\theta_{i}}\left(B_{2i}\cos\left(2\phi_{i}\right)-B_{1i}\sin\left(2\phi_{i}\right)\right) 
\end{align}
\end{subequations}
It should be noted that the ${\cal F}_{i}^{+,\times}$ coefficients are symmetric under the transformation $\beta\rightarrow -\beta$ and $\lambda\rightarrow \lambda+180^{\circ}$ for the gravitational wave source sky location, corresponding to the antipodal point on the sphere. Unfortunately, this degeneracy is manifest from the starting equations and will always result in a strong null signal at the antipodal point.

\section{Constructing a pulsar null stream}\label{sec.pns}

A null stream is constructed from the timing residuals of three
pulsars by noting that the same source polarization amplitudes,
${\tilde h}_{+,\times}(f)$, appear in the data stream from each
pulsar.  This fact is exploited by taking linear combinations of
pulsar data streams and factoring them in terms of the ${\tilde
h}_{+,\times}(f)$.  The \textit{null stream}, ${\tilde \eta}(f)$ is
the the linear combination of signals from the requisite set of
pulsars for which the putative gravitational wave signal vanishes.  If
the Fourier transform of the pulsar residuals is written as $ {\tilde
R}_{i}(f)$, then the null stream may be written as:
\begin{equation}
    {\tilde \eta} = \alpha_{1}\cdot{\tilde R}_{1}(f) +
    \alpha_{2}\cdot{\tilde R}_{2}(f) + \alpha_{3}\cdot{\tilde
    R}_{3}(f) = 0 \ ,
    \label{nullEta}
\end{equation}
where the $\alpha_{i}$ are linear coefficients. This factors into
\begin{eqnarray}
    {\tilde \eta} & = & {\tilde h}_{+}(f)\left[\alpha_{1}{\cal F}_{1}^{+}
    + \alpha_{2}{\cal F}_{2}^{+} + \alpha_{3}{\cal F}_{3}^{+}\right] 
    \nonumber \\
    & + & {\tilde h}_{\times}(f)\left[\alpha_{1}{\cal F}_{1}^{\times} +
    \alpha_{2}{\cal F}_{2}^{\times} + \alpha_{3}{\cal
    F}_{3}^{\times}\right]\ .
    \label{factoredEta}
\end{eqnarray}
The only way for ${\tilde \eta}(f) = 0$ generically is if the
coefficients multiplying ${\tilde h}_{+,\times}(f)$ in Eq.\
\ref{factoredEta} are zero:
\begin{eqnarray}
    \alpha_{1}{\cal F}_{1}^{+}
    + \alpha_{2}{\cal F}_{2}^{+} + \alpha_{3}{\cal F}_{3}^{+} & = & 0 
    \nonumber \\
    \alpha_{1}{\cal F}_{1}^{\times} +
    \alpha_{2}{\cal F}_{2}^{\times} + \alpha_{3}{\cal
    F}_{3}^{\times} & = & 0 \ .
    \label{eqn.nullCoefficients}
\end{eqnarray}
Setting the null stream to zero implies a relationship between the
arbitrary coefficients and the response functions. The system of equations is underdetermined for the $\alpha_i$, however the choice of three residual signals is crucial. If only two were chosen then only $\alpha_1=\alpha_2=0$ would solve the equations. Here we have a freedom to choose $\alpha_3$, but there is an obvious choice,
\begin{subequations}
\begin{align} \label{Coefficients}
    \alpha_{1}&=\mathcal{F}_{2}^{+}\mathcal{F}_{3}^{\times}-\mathcal{F}_{3}^{+}\mathcal{F}_{2}^{\times}\\
    \alpha_{2}&=\mathcal{F}_{3}^{+}\mathcal{F}_{1}^{\times}-\mathcal{F}_{1}^{+}\mathcal{F}_{3}^{\times}\\
    \alpha_{3}&=\mathcal{F}_{1}^{+}\mathcal{F}_{2}^{\times}-\mathcal{F}_{2}^{+}\mathcal{F}_{1}^{\times}.
\end{align}
\end{subequations}
Examination of the pulsar beam patterns ${\cal F}^{+,\times}$ in 
Eqs.~(\ref{BeamPattern1}) shows that these combinations have two free parameters: 
$\beta$ and $\lambda$, the position of the gravitational wave source 
on the sky. 

With the pulsar pattern functions ${\cal F}_{i}^{+,\times}$ in hand,
the derivation of the null stream ${\tilde \eta}(f)$ is reduced to
determining the values of $\{\alpha_{1},\alpha_{2},\alpha_{3}\}$ that
satisfy Eq.\ \ref{nullEta}.  The $\alpha_{i}$'s are combinations
of the pattern functions, which are themselves only functions of the
sky angles.  Operationally, the null stream search for the sky
location is then reduced to a two parameter minimization problem --
what values of the sky angles minimize ${\tilde \eta}(f)$?  In noise
free data, it will be a true nulling, by definition; in the presence
of noise the null stream will simply change the character of the
spectrum by suppressing features that are related to the gravitational
wave signal. 

These linear combinations can be constructed from any set of three
pulsar data streams and
minimized to determine the location of the source on the sky.  Given 
that modern pulsar timing arrays have greater than $50$ pulsars as part of 
the array, there are many different sub-arrays of pulsars that can be 
chosen to implement the pulsar null stream. This ability to chose 
sub-arrays can be exploited to increase the pointing ability of the 
technique.

Note that we have chosen above to do the analysis on a continuous source in the frequency domain, however, this need not be the case. As in previous work on null signals, \cite{TintoGursel,ZSS1,ZSS2,NullNetwork}, the analysis should work just as well for burst sources and in the time domain. The in depth analysis of combining various sub-arrays of pulsars has been favored over a full treatment of burst sources which will be treated in future work.

\section{Demonstration Using the Null Signal
Approach}\label{sec.demonstration}

\begin{figure}[H]
\begin{center}
\includegraphics[width=3.4in]{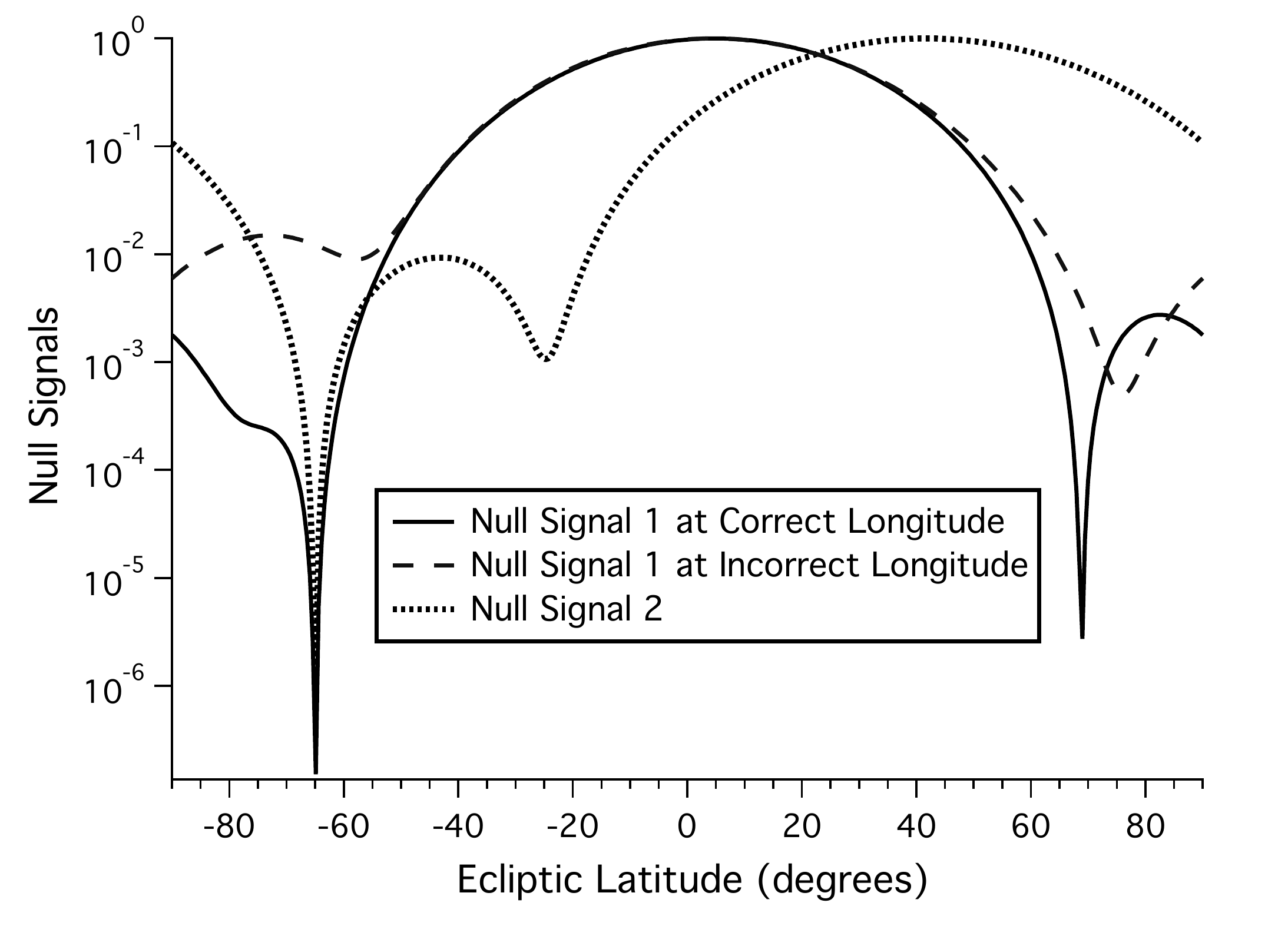}

\caption{The cross sections of the normalized null signal for three different cases is shown. The solid line is a cross section of $|\tilde{\eta}|^{2}$ with the value of $\lambda$ set to the ecliptic longitude of the source, $\lambda=195^{\circ}$. Notice that the strong dip at the correct latitude, $\beta=-64.7^\circ$. The dashed line is a cross section of the same null signal, but at a longitude that is not the correct longitude for the gravitational wave source, and hence does not have the dip. The dotted signal is another realization of the null signal, computed from an independent subarray made up of three other pulsars in the PTA. The cross section is taken at the correct longitude and one can see that even though the rest of the signal does not resemble the signal from the first three pulsars, it still possesses the same strong dip at $\beta=-64.7^{\circ}\!$.}
\label{Beta1}
\end{center}
\end{figure}

Using the null stream, Eq.~(\ref{nullEta}), we can localize the sky
position of a gravitational wave source by minimizing
$|\tilde{\eta}|^{2}$.  In the following example, a continuous, sinusoidal SMBH was modeled, using (\ref{Residual}), in Maple
with a sky position at {\small($\lambda=195^{\circ}$,
$\beta=-64.7^{\circ}$)}. We use the FFT of the residual signal and
the sky position of the pulsars to calculate the magnitude of
Eq.~(\ref{factoredEta}) as a function of the sky position of the
source.  Fig.~(\ref{Beta1}) shows cross sections of
$|\eta|^{2}$ along different longitudes.

\begin{figure}[htbp]
\begin{center}
\includegraphics[width=3.4in]{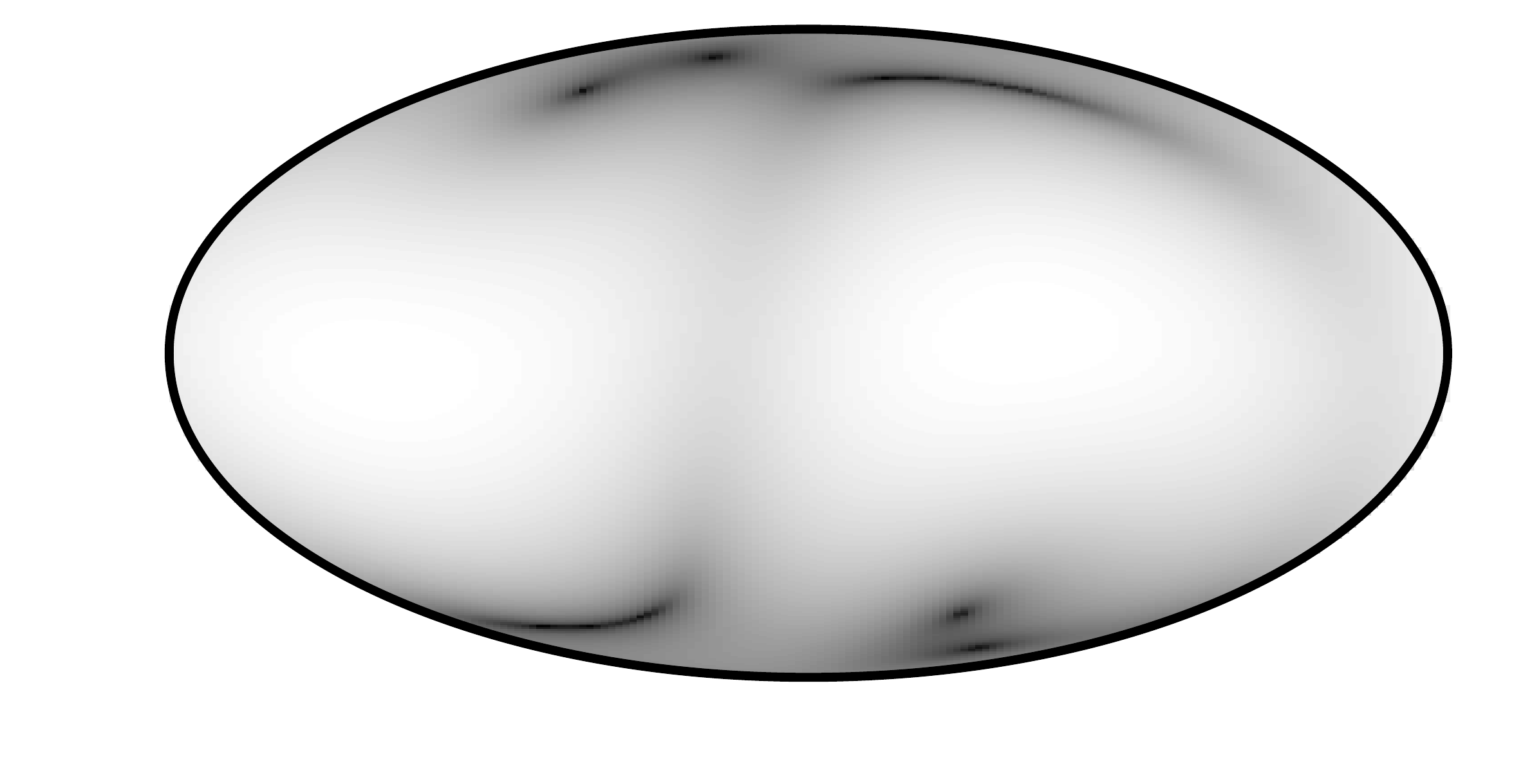}

\caption{The null signal, $|\tilde{\eta}|^{2}$, for 1 set of three pulsars is shown as a density plot. Darker areas represent low points in the signal, while lighter shades have a higher signal. Notice that there are deep nulls at the correct sky position ($\lambda=195^\circ$ and $\beta=-64.7^\circ$), the antipodal point, and other parts of the sky.}
\label{Density1}
\end{center}
\end{figure}

Notice the large dip at the correct value of the sky
position.  There is a secondary minimum in the cross section,
but it is an order of magnitude larger than the primary minimum.
Figure~\ref{Density1} shows the full sky density plot of the null signal.

\section{Error Versus Pulsar Separation}\label{sec.PulsarSep}

In a given null signal there are local minima which do not coincide with the correct sky position. This is evident in Fig.~\ref{Beta1}, where the null signal given by the solid line has a strong local minimum at around $75^{\circ}$. In fact, it is common, for a single sub-array of 3 pulsars, to have the absolute minimum for a null signal to be at an incorrect sky position.  The true null can be identified, and the overall size of the localization uncertainty can be minimized, by combining the null-signal from multiple sub-arrays of pulsars. This is one distinct advantage this method has over interferometric networks -- the large number of pulsars in the time array yields a large number of sub-arrays that can be combined to create a good null-stream pointing. In this section we will focus on how to characterize the errors in one sub-array. 

\begin{figure}[h]
\begin{center}
\includegraphics[width=3.5in]{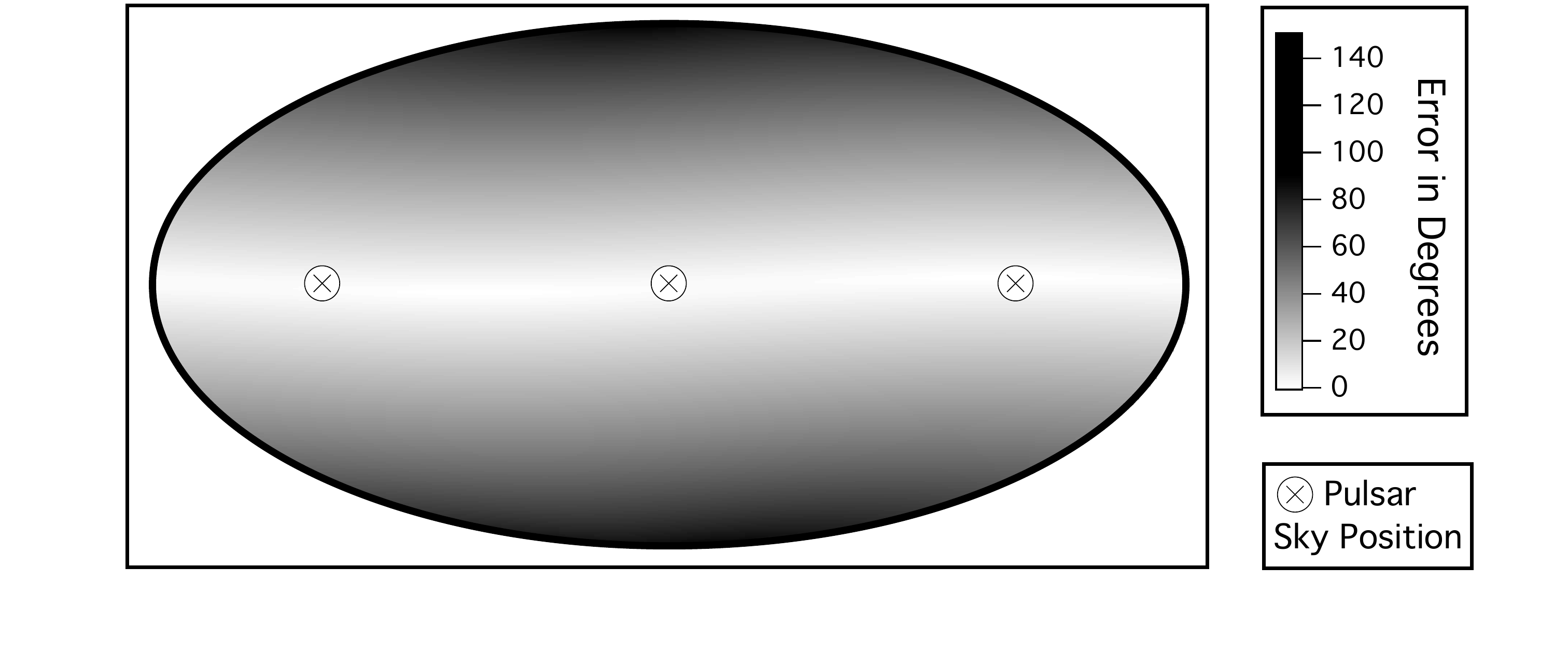}
\includegraphics[width=3.5in]{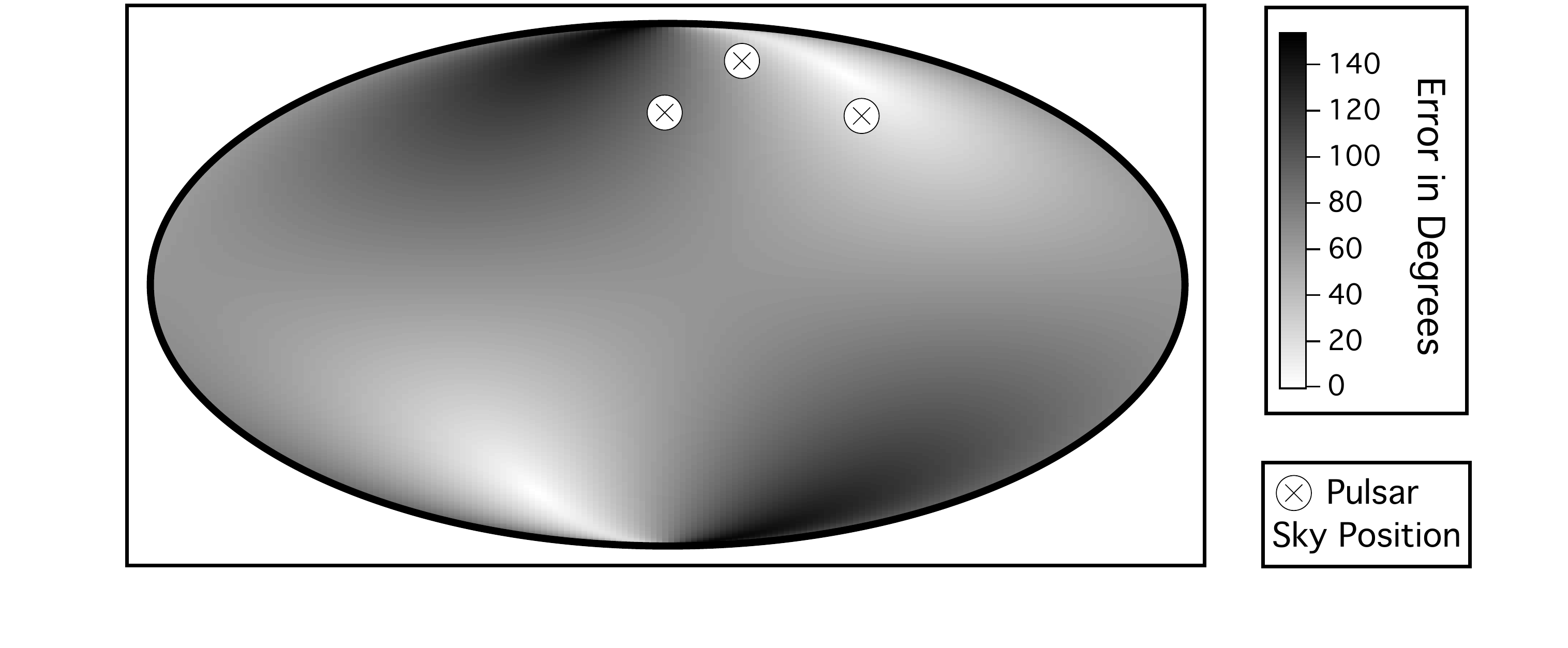}
\caption{The angular distance from the correct sky location is plotted as a function of the sky location for the given three pulsars. The darker areas represent areas where the null signal minimum gives a large error for the sky position of the gravitational wave source. The error in the sky position of the gravitational wave signal depends on both the angular distance of the sources from the pulsars and the pulsars angular distances from each other. }
\label{ErrorSkyMap}
\end{center}
\end{figure}

\begin{figure}[h]
\begin{center}
\includegraphics[width=2.5in]{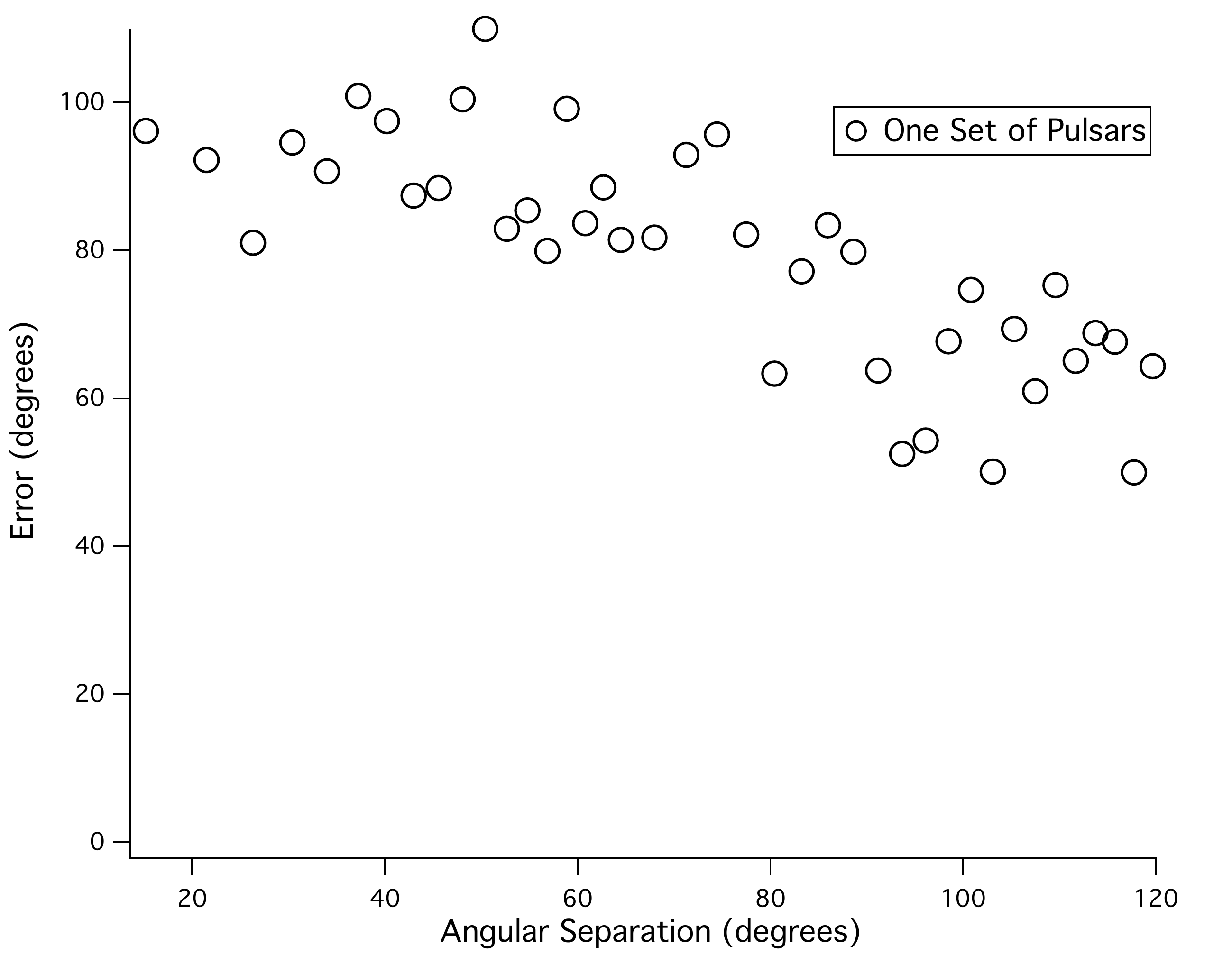}

\caption{Each data point is the error in sky position for an equilateral triangle of pulsars of given angular separation, averaged over 40 such triangles. The calculation was done using the same gravitational wave location used in previous examples, but randomly assigning the sky location of the pulsars. It is easy to notice the overall decrease in error as the separation increases to $120^{\circ}$. It should also be noted that these errors are still substantial even at larger separations.}
\label{ErrorTriangle1}
\end{center}
\end{figure}

The null stream localization error for a given sub-array of pulsars can be characterized by calculating the distance from the correct sky position to the absolute minimum for a given null signal. As one might suspect from localization schemes for other gravitational wave detectors as the angular distance between the pulsars in the null signal increases,  the error of the sky location decreases. In Fig.~\ref{ErrorSkyMap} are two examples of what the errors look like for a given null signal. These plots are constructed by putting three pulsars down at the locations indicated. Then for every point in the sky, we inject a signal, and ask where the minimum in the null signal is. The shading in Fig.~\ref{ErrorSkyMap} indicates the angular distance (error) from that point in the sky to the absolute minimum of the null signal.
Qualitatively it can be seen that there is less error when the pulsars in the null signal are more separated on the sky. This can be investigated more quantitatively by averaging over many sets of pulsars with the same separation. In Fig.~\ref{ErrorTriangle1} is a statistical analysis of sets of pulsars separated as equilateral triangles in the sky. Each data point represents the average of 40 different triangles sprinkled randomly across the sky. 
The same gravitational wave source sky location was used for all of the different pulsar placements. While the error to the minimum decreases substantially for a single sub-array, the absolute minimum still has a significant error, even at large angular separations.

\section{Multiple Sub-Arrays of Pulsars}\label{sec.MultSubArray}

While there is a strong null in the density map of $|\tilde{\eta}|^{2}$ shown
in Fig.~\ref{Density1}, there are strong secondary minima as well
across the entire sky. The
location and strength of the secondary minima are dependent on the
combined geometric orientations of the source and
the pulsars used in constructing the null stream.  Because the
strength of secondary minima varies dramatically with source position,
searches for sky position would benefit from reduction in size of the
secondary minima with respect to the null at the true sky location of
the source.  These secondary minima can be reduced, amplifying the
true null, by combining multiple null streams together.

In the case of gravitational wave interferometers we are limited by
the number of observatories at our disposal.  However, the PTA
catalog has more than $50$ pulsars. Any three can be used to construct a
null stream $\tilde \eta$.  We call any choice of three pulsars a
``sub-array.''  The sub-array null signals are combined into a single signal by taking their product. This acts to strengthen the signal while suppressing the random fluctuations in the data. For purposes of comparison the individual null streams are all normalized by dividing a null signal by its maximum before taking their product.
\begin{figure}[htbp]
\begin{center}
\includegraphics[width=3.4in]{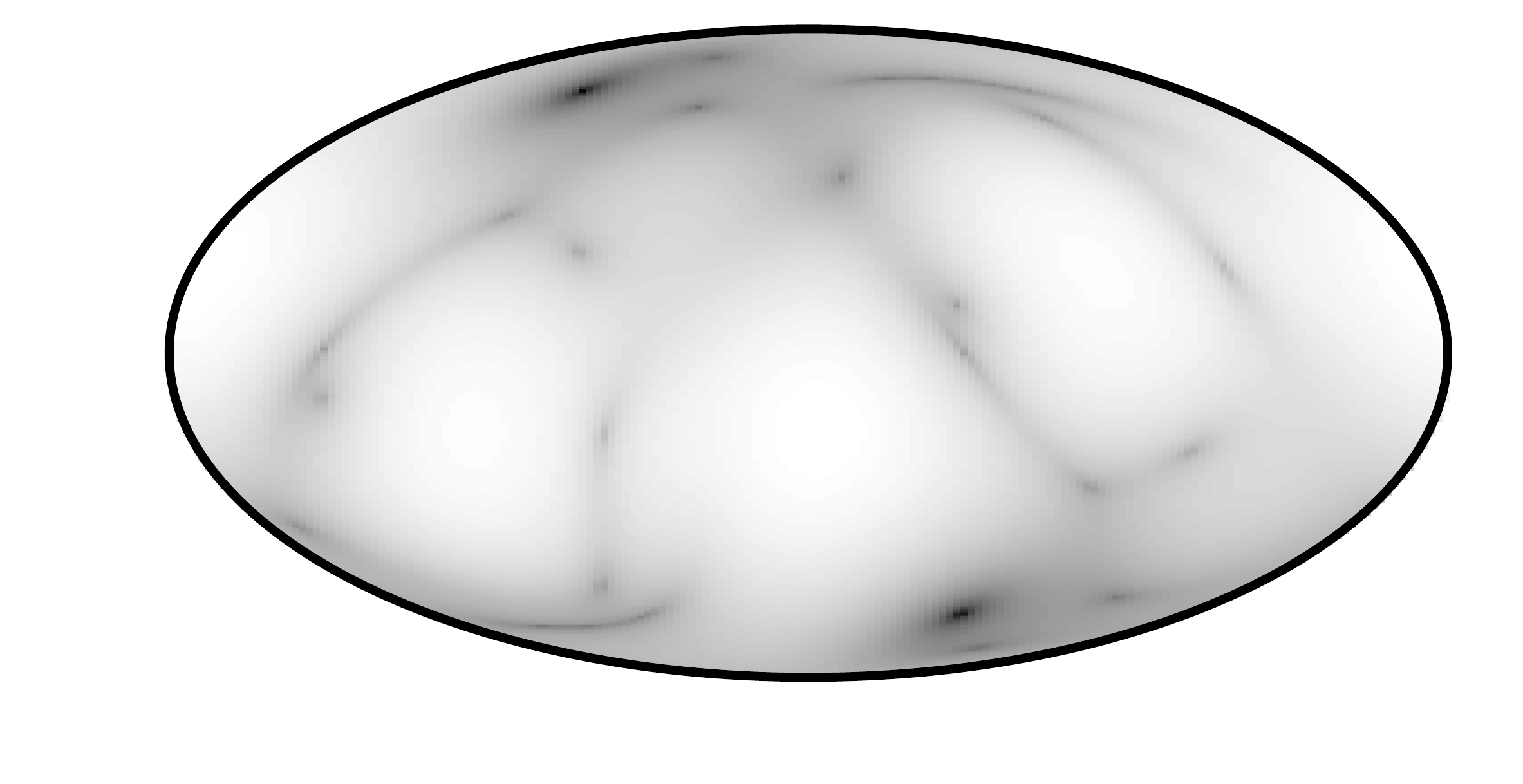}
\includegraphics[width=3.4in]{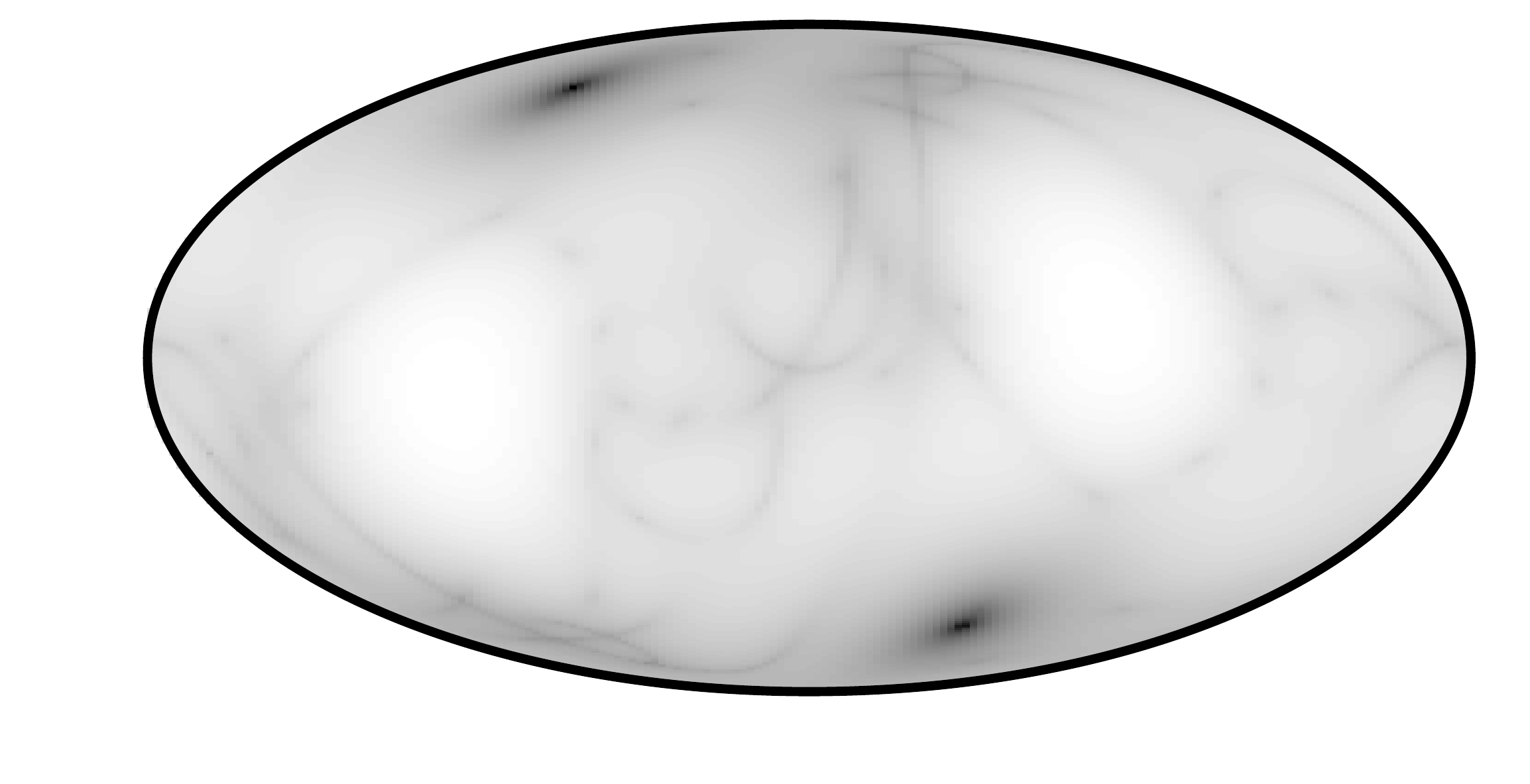}
\caption{Density plot of a null signal product using 3 sets of pulsars (top) and 9 sets of pulsars (bottom).}
\label{DensitySets}
\end{center}
\end{figure}

\begin{figure}[htbp]
\begin{center}
\includegraphics[width=3.5in]{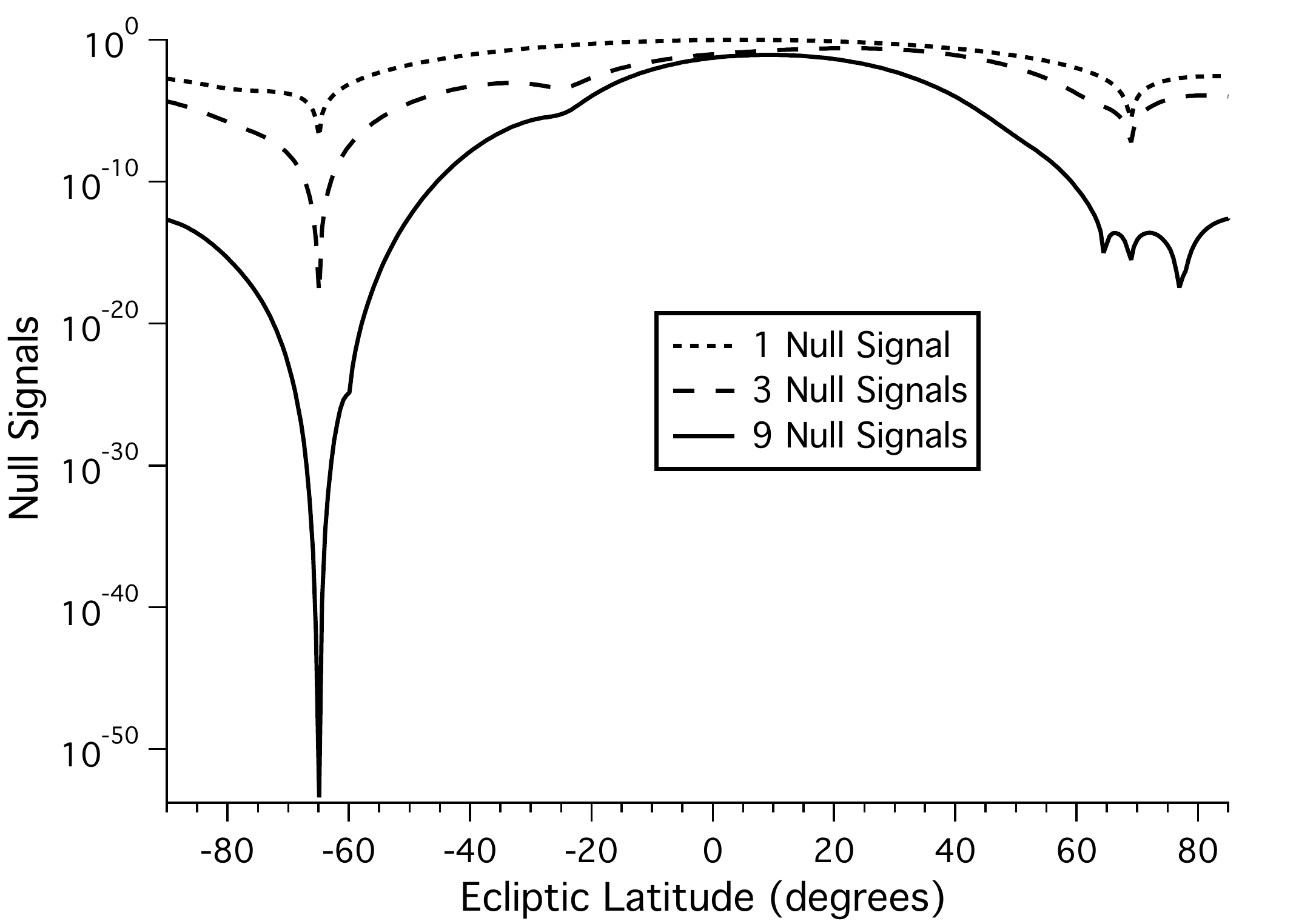}

\caption{Null signals using1, 3 and 9 sets of pulsars with the value of
$\lambda=195^{\circ}$.}
\label{BetaNulls}
\end{center}
\end{figure}

Combining multiple null streams from independent sub-arrays
immediately suppresses the secondary minima, quickly revealing the
location of the null at the true sky location of a source.
The top sky map in Fig.~\ref{DensitySets} shows the product of 3 independent sets of pulsars. There are still swaths of low $\eta$ values where the 
secondary minima are seen in Fig.~\ref{Density1}, but their relative 
strength is greatly reduced. The lower sky map in Fig.~\ref{DensitySets} shows the product of  9 independent null signals, revealing the location of the true null 
on the sky.   
Fig.~\ref{BetaNulls} shows the one dimensional cross-sections through the 
parameter space at a constant value of $\lambda$, showing the strong null at the true source location, 
and only minor variations across the rest of the sky.
\begin{figure}[htbp]
\begin{center}
\includegraphics[width=3.5in]{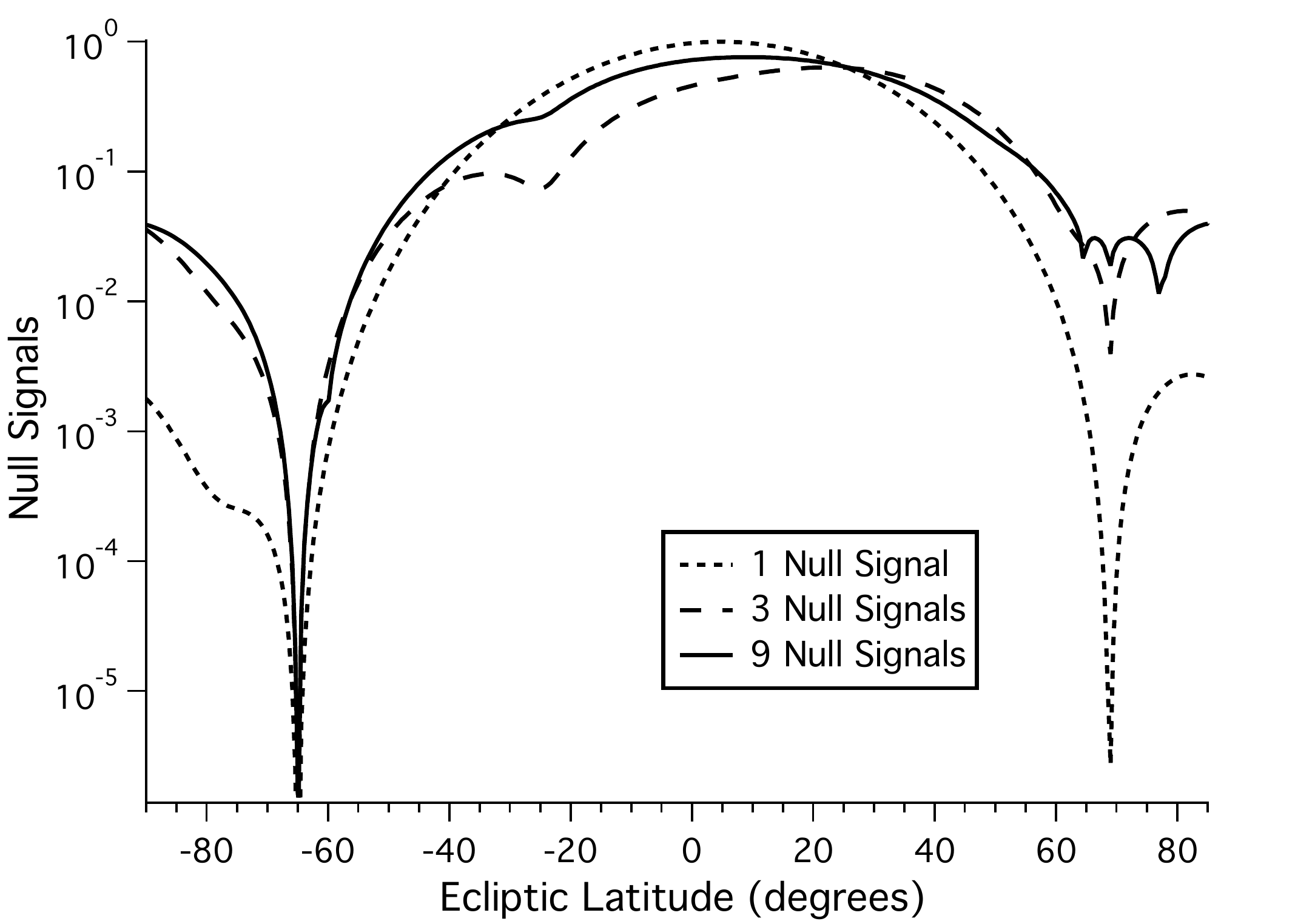}

\caption{This data is identical to Fig~\ref{BetaNulls}, but the $n$th root has been taken, where $n$ is the number of sub-arrays in the null stream. The $n$th root is taken in localization comparisons so that the width of the minima can be compared over roughly the same range of values of the null signal.}
\label{BetaNullsRoots}
\end{center}
\end{figure}

\begin{figure}[htbp]
\begin{center}
\includegraphics[width=3.3in]{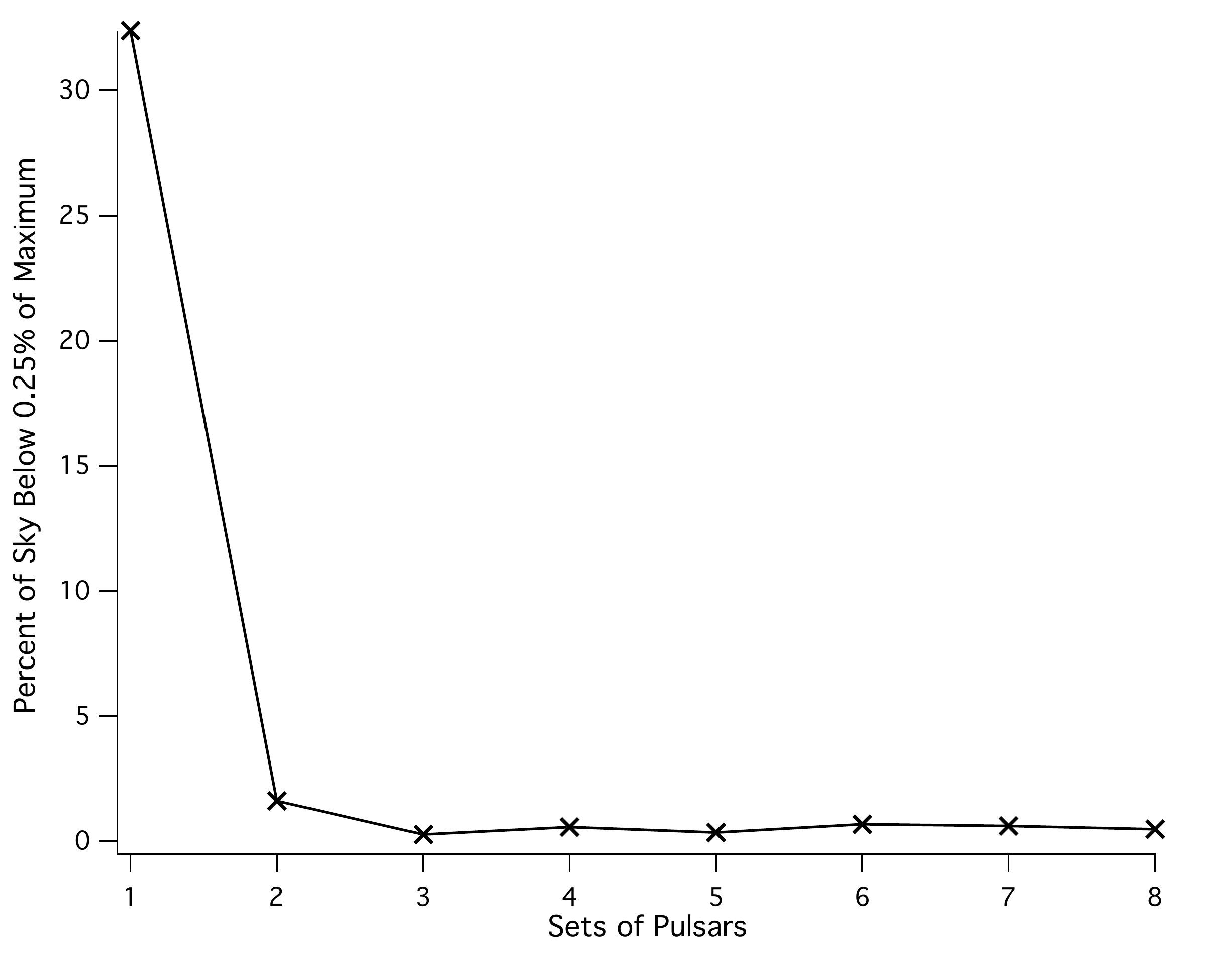}

\caption{This figure shows the percentage of the sky that falls under
a given cutoff value of the null signal.  The localization becomes
precise very quickly as we increase the number of sets of pulsars.
For this example, after 3 sets of pulsars the localization does not increase
significantly.  The cutoff is given as a percentage of the maximum,
and the percentages are normalized for comparison because of the
different powers for the multiple sub-array null signals.}
\label{LocalizationNumberSets}
\end{center}
\end{figure}

\begin{figure}[htbp]
\begin{center}
\includegraphics[width=3.3in]{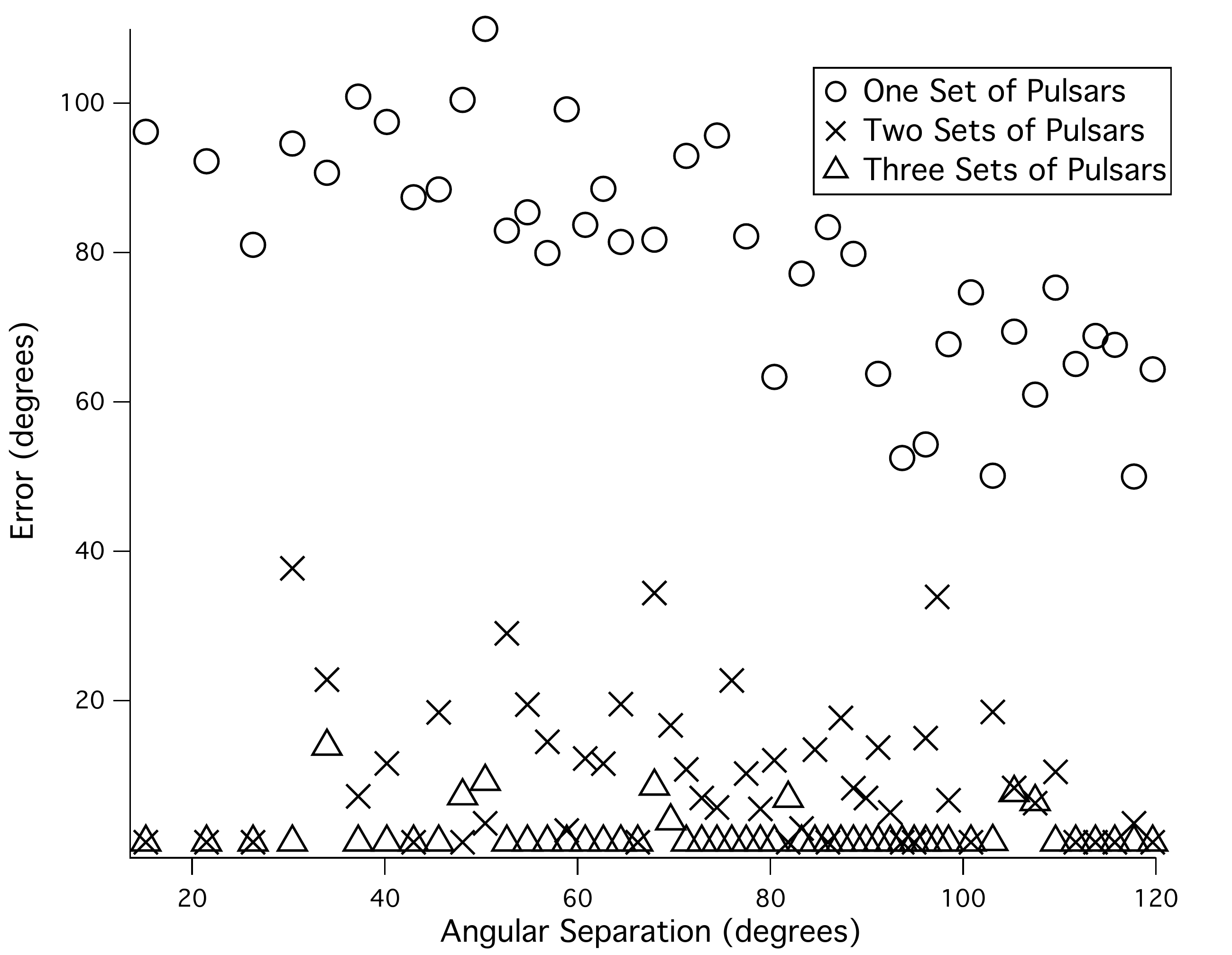}

\caption{Each data point is the error in sky position for an equilateral triangle of pulsars of given angular separation, averaged over 40 such triangles. The calculation was done using the same gravitational wave location used in previous examples, but randomly assigning the sky location of the pulsars.}
\label{ErrorArea}
\end{center}
\end{figure}
While there are immediate gains in localization ability by combining
multiple sub-arrays, relative gains are eventually reduced each time a new sub-array is added to the product. Figure~\ref{LocalizationNumberSets} shows one example of how localization
ability grows with the number of sub-arrays.  

To quantify the amount
of localization from multiplying multiple null signals we first normalize the signals further by taking the $n^\text{th}$ root of the signal, where $n$ is the number of null signals in the product (see Figure~\ref{BetaNullsRoots} to see what this looks like). This gives the null signals the same range. Then define a
cutoff value of the null signal given as a percentage of the maximum
of the null signal.  Looking at the number of square degrees that
falls under this cutoff gives a measure of how much one can localize
the position of a given source.

These localizations vary depending on the relationship between the gravitational wave source and the pulsars. Another way to characterize the localization is to repeat the more statistical analysis done using sets of equilateral triangles to characterize the error of the minimal signal for a given null signal. Fig.~\ref{ErrorArea} shows how the product of null signals significantly decreases the errors in the minimum. Once we get past 3 sub-arrays the error is significantly reduced.

\section{Noise and Null Streams}\label{sec.noise}

The formal presentation of the null stream solution given by Eq.\ 
\ref{nullEta} is an idealized, mathematical observation about the 
nature of the pulsar timing data and the putative signals it 
contains. Real data, however, is subject to the presence of noise 
from random processes that can reduce the promising capabilities of 
this localization method.

The effects of noise can be considered in this demonstration by 
injecting noise into the residual data for the pulsars at various 
levels, then examining how it affects the solutions for sky positions.

To produce a noisy data set, white gaussian noise $n_{i}(t)$ is 
generated using Maple and added to the residuals $R_{i}(t)$ for each pulsar. In the examples here the mean of the noise is set to zero while the standard deviation is set to the maximum amplitude of the gravitational wave source. While an SNR$\;\sim8$ is an accepted standard for detection, here it is difficult to see the effect that noise has on the null signal until the noise is closer to SNR$\;\sim1$.

\begin{figure}[htbp]
\begin{center}
\includegraphics[width=3.5in]{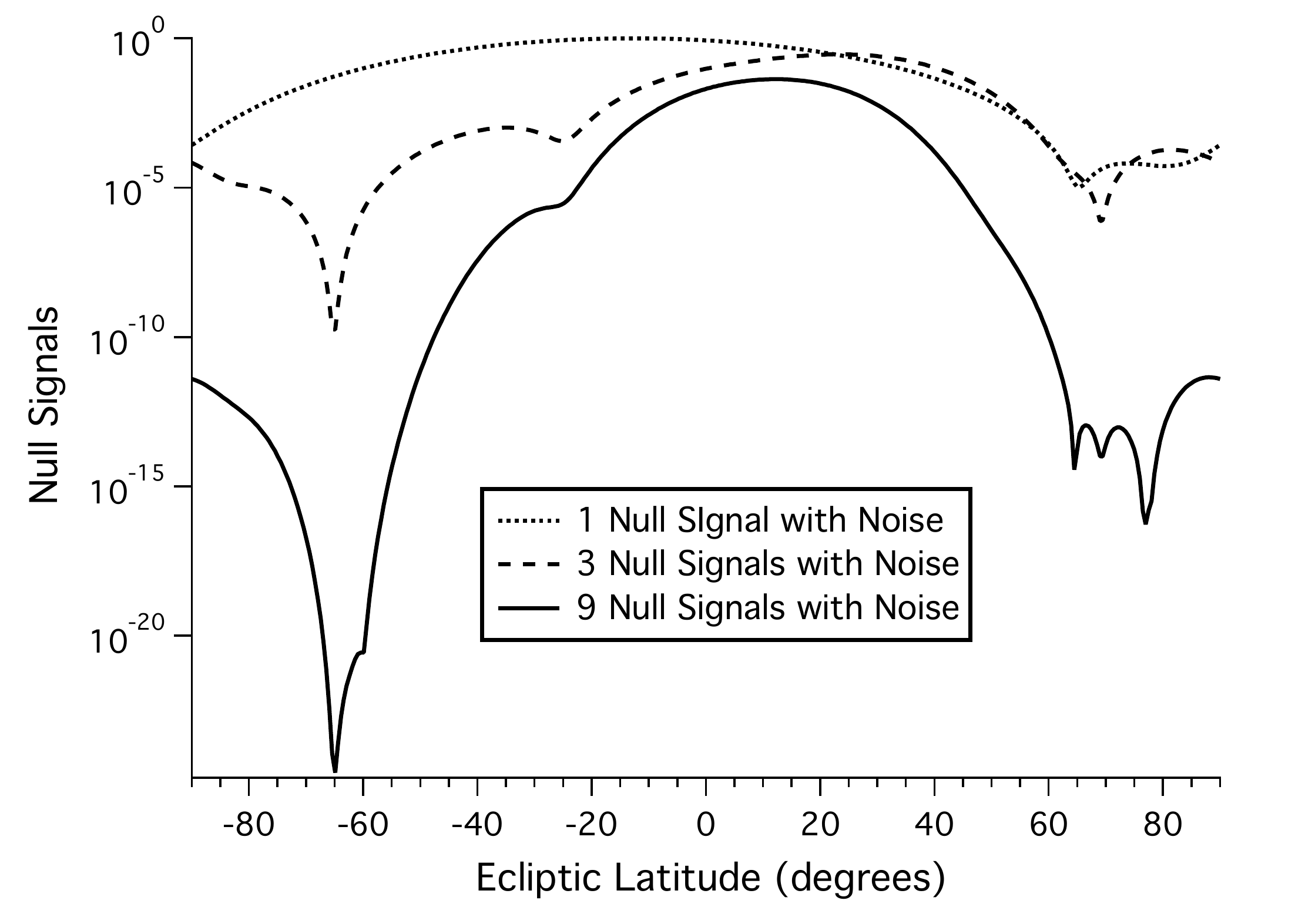}

\caption{Null signals with white gaussian noise, SNR$\;\sim1$, added using 1, 3 and 9 sub-arrays of pulsars. As before these are cross-sections through $\lambda=195^{\circ}$, the correct ecliptic longitude for the source. Notice that with only one noisy signal it is difficult to discern the correct latitude for the source, however with a product of three null signals we see a marked dip, and with a product of nine null signals an even more localized minimum.}
\label{BetaNullsNoise}
\end{center}
\end{figure}

\begin{figure}[htbp]
\begin{center}
\includegraphics[width=3.4in]{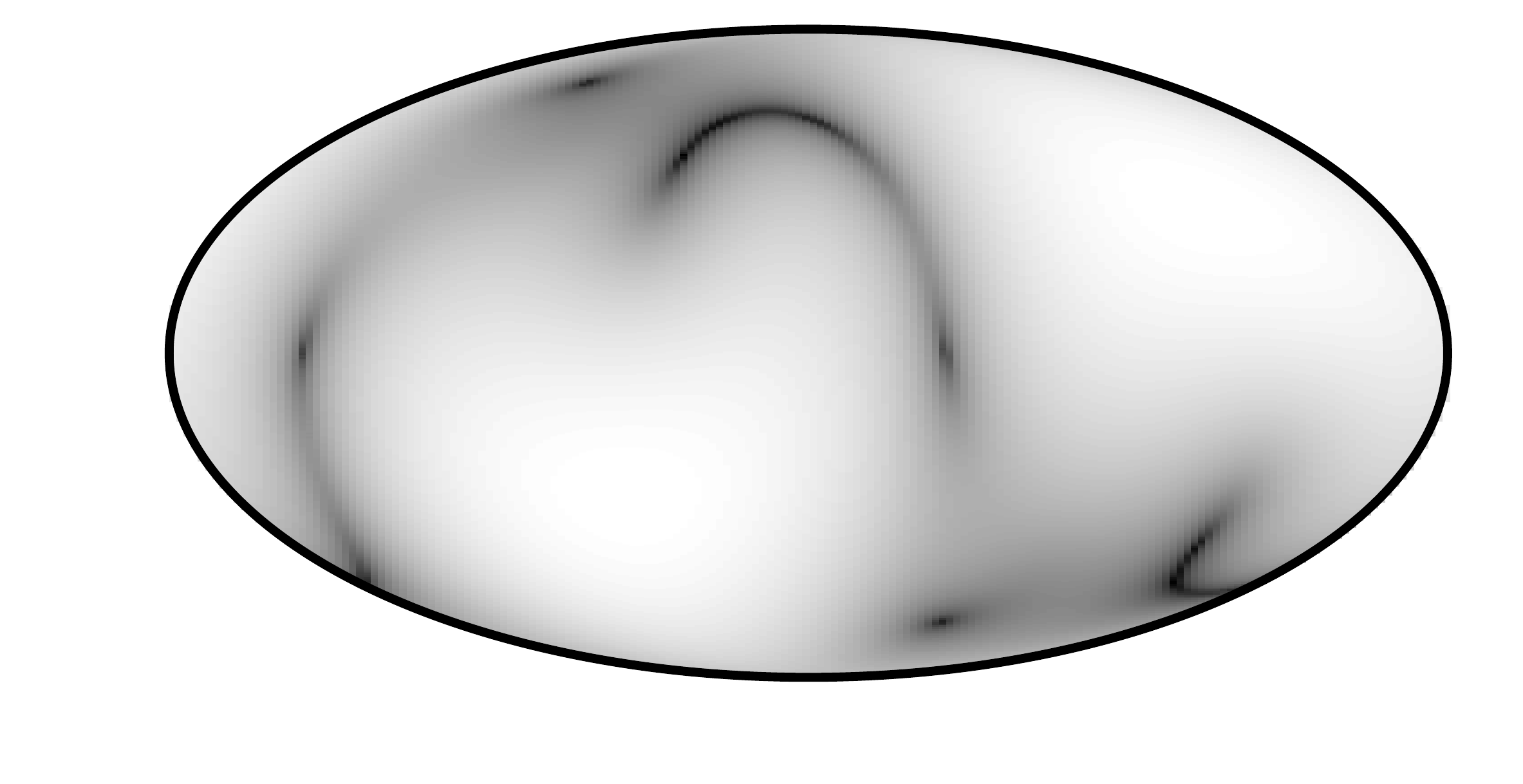}
\includegraphics[width=3.4in]{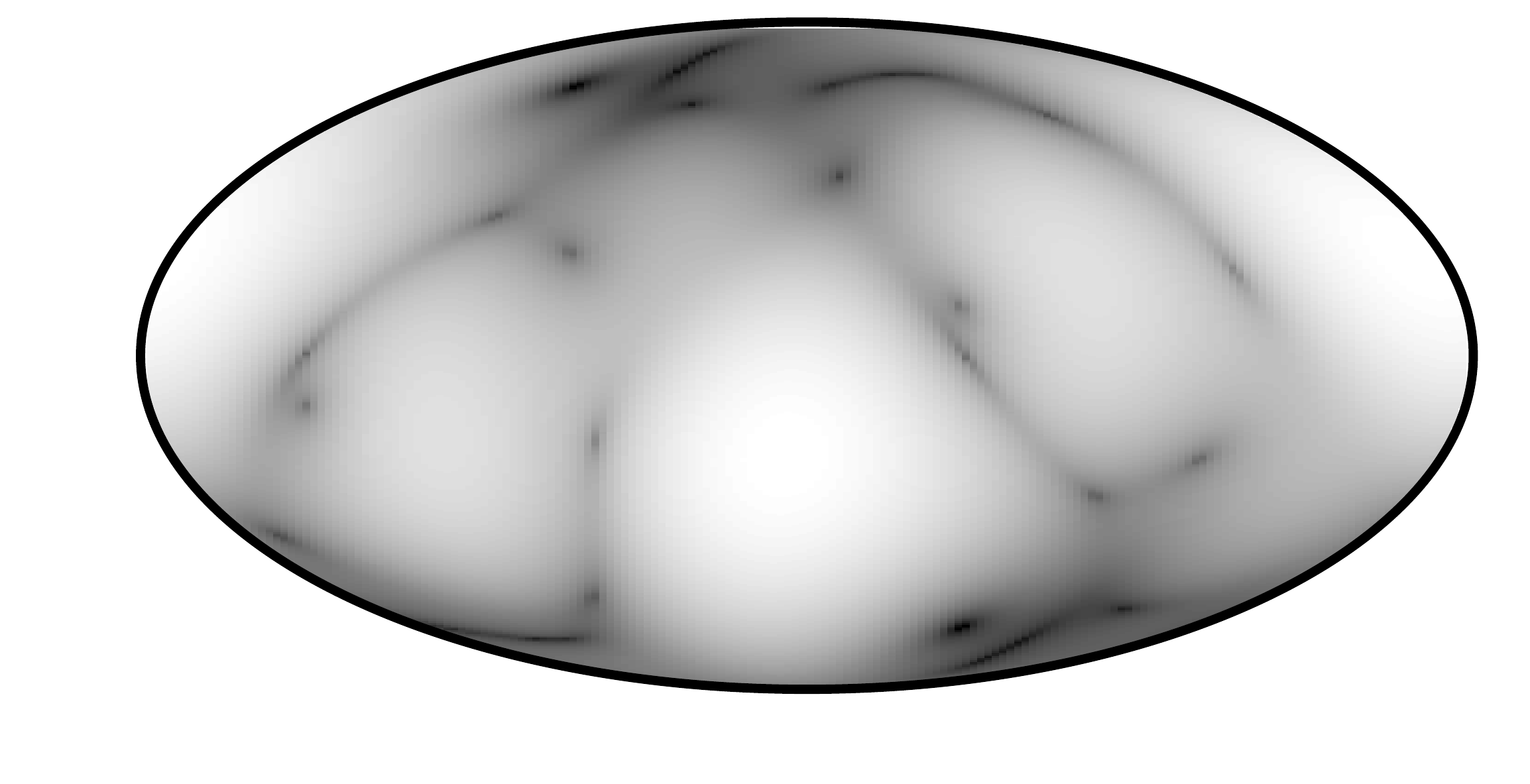}
\caption{These density plots are sky maps of a null signal with white gaussian noise, SNR$\;\sim1$. The top figure is a single null signal and resembles Fig.~\ref{Density1} strongly, however there is no dip at the correct sky location, as can be seen in the crossection in Fig.~\ref{BetaNullsNoise}. The lower figure is a product of three null signals, where we see that we have regained the strong localization.}
\label{DensitySetsNoise}
\end{center}
\end{figure}

In Figures~\ref{BetaNullsNoise} and \ref{DensitySetsNoise} we see that the added noise affects the null signal, but fairly subtly. In Fig.~\ref{BetaNullsNoise} we see that by the time we have the product of three null signals we have regained much of the localization we had without noise. 

Perhaps counter-intuitively, the null stream approach still provides 
good localization even when the noise is comparable to the size of 
the signal, a situation that would be unfathomable in traditional 
parameter estimation. This can be understood by considering that the 
product of null signals amplifies the null (by making the null smaller), while white noise is just as likely to increase the signal as decrease it.

\section{Discussion}\label{sec.discussion}
Null stream mapping of gravitational wave sources 
relies on the fact that there are correlated gravitational wave
signals between detectors.  The underlying premise of the null stream
construction is that for a collection of pulsars observing the same
source, the gravitational wave signal is common to all pulsars in the
array, but modified by geometric factors related to the relative
position of the source on the sky.  We have shown how a linear combination of three pulsar timing streams gives a signal that is minimized at the correct sky location of the gravitational wave source, though not as localized as one might need for electromagnetic counterpart searches. Further we have characterized the error and localization ability of one null signal. Though there are significant errors with one null signal when multiple signals are combined as products the errors decrease and the localization increases dramatically. 

The techniques here have focused on analysis in the frequency domain specialized for looking at stable sinusoidal signals expected from super massive black hole mergers, however, future work will consider how these sky localization techniques will work for burst type sources in the time domain.

\acknowledgments SLL acknowledges support from National Science
Foundation award PHY-0970152, and from NASA award NNX13AM10G.


\end{document}